\documentclass[a4paper,11pt]{article}
\usepackage{amsmath,amssymb,amsthm,epic,eepicemu,graphicx}
\usepackage[english]{babel}
\usepackage[pdfpagemode=UseOutlines,bookmarksnumbered=true,,bookmarksopen=true,
 pdfhighlight=/P,colorlinks=true,linkcolor=blue,citecolor=blue,urlcolor=blue,
 unicode=true]{hyperref}
\usepackage[width=0.9\textwidth,indention=0mm,justification=centerlast,
 font=small,labelfont=bf,labelsep=period]{caption}

\title{Cauchy problem for integrable discrete equations on quad-graphs
\bigskip\\
\hfill\parbox{6cm}{\begin{center}\normalsize\em
 Dedicated to S.P.~Novikov\\
 on his 65 birthday
\end{center}}}

\author{
V.E.~Adler\thanks{Institut f\"ur Mathematik, Technische Universit\"at Berlin,
Str.~des 17.~Juni 136, 10623 Berlin, Germany. On leave from Landau Institute
for Theoretical Physics, Chernogolovka, Russia. E-mail: {\tt adler@itp.ac.ru}}
\and
A.P.~Veselov\thanks{Loughborough University, Loughborough, Leicestershire LE11
3TU, UK and Moscow State University, Russia. E-Mail: {\tt
A.P.Veselov@lboro.ac.uk}}}

\date{\empty} 

\newcommand{\Integer}{{\mathbb Z}}
\newcommand{\Complex}{{\mathbb C}}
\newcommand{\CP}{{\mathbb CP}}
\newcommand{\ind}{\mathop{\rm ind}}
\def\bfput#1{{\thicklines
     \put(-0.4,-0.4){#1}\put(-0.4, 0.4){#1}
     \put( 0.4,-0.4){#1}\put(-0.4,-0.4){#1} }}

\theoremstyle{plain}
\newtheorem{theorem}{Theorem}
\theoremstyle{definition}
\newtheorem{definition}{Definition}

\newenvironment{Proof}[1][]
{\begin{proof}}
{\end{proof}}

\begin{document}
\maketitle
\begin{abstract}
Initial value problems for the integrable discrete equations on quad-graphs are
investigated. We give a geometric criterion of when such a problem is
well-posed. In the basic example of the discrete KdV equation an effective
integration scheme based on the matrix factorization problem is proposed and
the interaction of the solutions with the localized defects in the regular
square lattice are discussed in details. The examples of kinks and solitons on
various quad-graphs, including quasiperiodic tilings, are presented.
\end{abstract}

\section{Introduction}\label{s:intro}

The discrete potential KdV (dpKdV) equation
\begin{equation}\label{dpKdVmn}
 (v_{m+1,n+1}-v_{m,n})(v_{m+1,n}-v_{m,n+1})=\alpha_m-\beta_n,\quad
 m,n\in\Integer
\end{equation}
can be considered as a simplest representative of the integrable nonlinear
discrete equations in two dimensions. This field is widely studied and many
other examples can be found in the literature, see e.g.
\cite{Bi,H,M,QNCL,PNC,CNP,NC}. However, we restrict ourselves to this simplest
model, since our aim here is to analyze the generalizations of another kind.

More precisely, we will discuss the role of the {\em support} of the discrete
equations, that is, roughly speaking, the set where the independent variables
live. Of course, the most natural way to discretize 2-dimensional PDE is to
consider equations on the regular square grid $\Integer^2$ as above, but recent
results \cite{A,BS,NW,Ni,ABS,ND,DN,DN02} demonstrate that possibly more general
synonym of `2D' in the discrete case is `planar graphs'.

It should be noted that non-standard planar graphs were considered in
Mathematical Physics already quite a while ago (see e.g. Korepin's works
\cite{Kor,Kor2} where the solvable spin models on the quasi-crystallic tilings
were investigated). A systematic theory of the linear difference operators on
the graphs in relation to soliton theory has been initiated by S.P.~Novikov
(see \cite{ND,No,DN,DN02,KN}).

We start in Section \ref{s:Qeq} with the necessary information about integrable
equations on {\em quad-graphs}, which are planar graphs with quadrilateral
faces. In this case the integrability can be understood as the so called 3D
consistency property, introduced and studied in \cite{BS,NW,Ni,ABS}. The
discrete  potential KdV equation on quad-graph reads locally exactly as on the
square lattice:
\begin{equation}\label{dpKdV}
 (v_{12}-v)(v_1-v_2)=\alpha_1-\alpha_2,
\end{equation}
but now it is assumed that the fields $v$ are assigned to the vertices of the
quad-graph and the parameters $\alpha$ are assigned to the edges, as shown on
the Fig.~\ref{fig:quad} below, and the relation (\ref{dpKdV}) must be fulfilled
for each face of the graph.

Section \ref{s:ivp} is devoted to the general discussion of the possible
settings of Cauchy problem or initial value problem (IVP) on quad-graphs. At
the first sight the question about the well-posedness of IVP is not related to
integrability but in fact it does very much (see e.g. Fig.~\ref{fig:defect10},
\ref{fig:aa} below). The 3D consistency property plays an essential role here.

Section \ref{s:corr} contains the main result of the paper: a criterion for the
existence and uniqueness of the solution of Cauchy problem for an integrable
equation on a quad-graph (Theorem \ref{th:corr}).

In section 5 we discuss the effect of interaction of the solutions of the dpKdV
equation with the localized defects in the regular square lattice. We introduce
a notion of the {\em weak defect} and show that such a defect does not affect
the dynamics outside it. An integration scheme based on the matrix
factorization problem plays a crucial role in the proof.

Several examples of the kink/solitons solutions on various quad-graphs,
including  quasiperiodic tilings are presented in the section \ref{s:kink}.

In the last section we discuss the solutions of the linear discrete wave
equation
\[
 v_{12}-v_1-v_2+v=0
\]
on quad-graphs. Surprisingly enough, in contrast to the non-linear case these
solutions have a non-trivial interaction with weak defects, although the effect
is limited (see Theorem~\ref{th:wave}).

\paragraph{Remark.} The first version of this paper published in {\em Acta
Applicandae Mathematicae \bf 84} (2004)
\href{http://dx.doi.org/10.1007/s10440-004-5557-9}{237--262}
contained some inaccuracies. The main issue concerned the notion of
solutions which were tacitly assumed to be nonsingular. This problem was
addressed in DSc thesis of V.A. (June 2010, electronic version in Russian is
available at \url{http://www.itp.ac.ru/~adler/dd.html}) where this assumption
was explicitly formulated. In the present version we made minimal corrections
in this line. For the first time, the notion of singular solutions was
introduced in \cite{ABS_2009} in the context of classification of 3D-consistent
equations. The study of general solutions should require a proper
algebro-geometric treatment of the Cauchy data in the spirit of Okamoto and
Sakai. The Cauchy problem on quad-graphs was also discussed 
recently by van der Kamp \cite{Kamp}, who noticed some deficiencies 
in the first version of our paper too.

\section{Quad-graph equations}\label{s:Qeq}

We start with the results of the paper \cite{ABS} where all the integrable
cases were classified among the certain class of discrete equations on
quad-graphs. The basic building block of such equation is an equation on the
elementary quadrilateral of the form
\begin{equation}\label{Qeq}
 Q(v,v_1,v_2,v_{12};\alpha_1,\alpha_2)=0,
\end{equation}
where the field variables $v\in\Complex$ are assigned to the vertices and the
parameters $\alpha\in\Complex$ are assigned to the edges of quadrilateral, as
shown on the Fig.~\ref{fig:quad}. The following properties are assumed to be
satisfied:

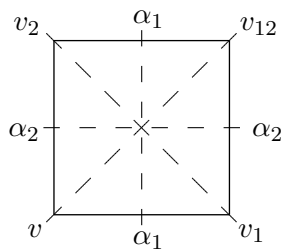
\begin{figure}[t]
\begin{center}\setlength{\unitlength}{0.06em}
\begin{picture}(200,150)(-50,-20)
 \path(0,0)(100,0)(100,100)(0,100)(0,0)
 \dashline{10}(-6,50)(106,50)
 \dashline{10}(50,-6)(50,106)
 \dashline{13}(-4,-4)(104,104)
 \dashline{13}(-4,104)(104,-4)
 \put(-23,106){$v_2$}  \put(106,106){$v_{12}$}
 \put(-15,-13){$v$}    \put(105,-13){$v_1$}
 \put(-26, 45){$\alpha_2$} \put(113, 45){$\alpha_2$}
 \put( 45,-17){$\alpha_1$} \put( 45,110){$\alpha_1$}
\end{picture}
\caption{An elementary quadrilateral and its $D_4$ symmetry}\label{fig:quad}
\end{center}
\end{figure}

(i) the parameters $\alpha$ on the opposite edges of any face are equal;

(ii) the equation (\ref{Qeq}) is invariant under the group $D_4$ of the square
symmetries:
\begin{multline}\label{Qsym}
 Q(v,v_1,v_2,v_{12};\alpha_1,\alpha_2)
  =\varepsilon Q(v,v_2,v_1,v_{12};\alpha_2,\alpha_1) \\
  =\sigma Q(v_1,v,v_{12},v_2;\alpha_1,\alpha_2),\quad
  \varepsilon,\sigma=\pm1;\qquad
\end{multline}

(iii) the function $Q$ is a polynomial of degree one with respect to each field
variable.

Let $\Gamma$ be a finite quad-graph and $V(\Gamma)$, $E(\Gamma)$, $F(\Gamma)$
be the sets of its vertices, edges and faces respectively. A mapping
$\alpha:E(\Gamma)\to\Complex$ which satisfies the property (i) is called {\em
labelling}. The choice of labelling allows us to define the {\em quad-graph
equation}
\begin{equation}\label{Q}
 Q(v_i,v_j,v_k,v_l;\alpha_{ij},\alpha_{ik})=0,\quad (i,j,l,k)\in F(\Gamma).
\end{equation}
Property (i) means that the values of parameters $\alpha_{ij}$ are constant
along any {\em strip} (or, {\em characteristic}), which is a sequence of
quadrilaterals adjacent by the opposite edges. Some characteristics are shown
at the pictures below by dashed lines crossing the adjacent edges of the faces
in the strip. We will see in the next sections that this notion is very
important in the theory of quad-graph equations and plays the role similar to
the characteristics for the second-order hyperbolic equations. In the
particular case of dpKdV equation on the square lattice (\ref{dpKdVmn}) we have
two families of the parameters $\alpha_m$ and $\beta_n$ assigned to the
vertical and horizontal straight strips, but in the general case the strips may
bend and intersect in a very complicated way.

Property (ii) allows us to define equation on each face independently on its
orientation. Property (iii) means that the equation (\ref{Qeq}) can be solved
with respect to a field at any vertex as a linear fractional expression of
other three fields. It may happen that both numerator and denominator vanish,
in which case the corresponding field is undetermined. It is rather difficult
either to control such solutions or to formulate conditions for quad-graph
which guarantee their absence. We solve this difficulty in a most radical way,
just by imposing the absence of singularities into the definition of the
solution.

\begin{definition}\label{def:sol}
A solution $v:V(\Gamma)\to\CP^1$ of equation (\ref{Q}) on the labelled
quad-graph $\Gamma$ is called {\em nonsingular} if $\partial Q/\partial
v_s\ne0$, $s=i,j,k,l$ for all faces $(i,j,l,k)\in F(\Gamma)$. Otherwise, we
call the solution {\em singular}.
\end{definition}

For example, for the dpKdV equation (\ref{dpKdV}) the solution is singular with
respect to $v_{12}$ if $v_1=v_2$. Moreover, if $\alpha_1\ne\alpha_2$ then
$v=\infty\in\CP^1$ and $v_{12}\in\Complex$ is arbitrary and if
$\alpha_1=\alpha_2$ then $v\in\Complex$ is arbitrary as well. For integrable
quad-equations from the list \cite{ABS} the singular solutions are in general
characterized by the system of two biquadratic equations \cite{ABS_2009}.

The requirement of nonsingularity leads to some additional restrictions for
the labelling of quad-graph and also for the quad-graph itself. We will say
that the labelling of the quad-graph is {\em proper}, if the corresponding
equation is irreducible for each face.

\begin{figure}[t]
\centerline{\includegraphics[width=100mm]{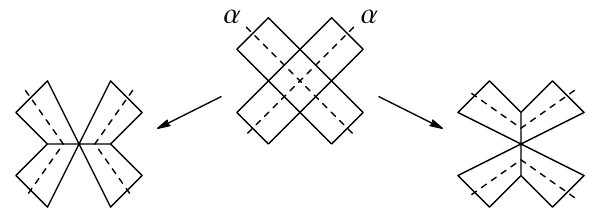}}
\caption{Eliminating of intersection of strips with the same parameter}
\label{fig:aa}
\end{figure}

In particular, it is easy to see that equations from the list \cite{ABS} become
reducible for some special choice of parameters. First of all, when
$\alpha_j=\alpha_i$, all equations with no exceptions are reduced to
\[
 (v-v_{12})(v_1-v_2)=0.
\]
All solutions of this equation are singular, therefore the necessary condition
for existence of nonsingular solution on the quad-graph is that the parameters
on intersecting edges of any face should be different. This is equivalent to
the prohibition of intersecting strips which carry the same parameter, in
particular, the self-intersecting strips. Note that this requirement can be
satisfied by eliminating the illegal faces as shown on fig. \ref{fig:aa}.
Clearly, the two ways to identify the opposite vertices correspond to the
choice between solutions $v=v_{12}$ and $v_1=v_2$ (it should be stressed that
this choice is made in advance, not in the process of the solution
construction).

We would like to mention that the absence of self-intersecting strips plays a
crucial role in the general problem of embedding of a quad-graph into cubic
lattice which was initiated by S.P. Novikov and investigated in details by
Dolbilin et al \cite{DS}. In fact our proof of Theorem \ref{th:corr} below is
based on a similar geometric consideration, but instead of cubic lattice we
consider a multidimensional cube.

Note that other equations may have different prohibited special values. For
instance, the equation ($Q^\delta_3$)
\[
\begin{aligned}
 & \Bigl(\alpha_i-\frac1{\alpha_i}\Bigr)(vv_i+v_jv_{ij})
  -\Bigl(\alpha_j-\frac1{\alpha_j}\Bigr)(vv_j+v_iv_{ij}) \\
 &-\Bigl(\frac{\alpha_i}{\alpha_j}-\frac{\alpha_j}{\alpha_i}\Bigr)(vv_{ij}+v_iv_j)
  -\frac{\delta}{4}\Bigl(\alpha_i-\frac1{\alpha_i}\Bigr)
   \Bigl(\alpha_j-\frac1{\alpha_j}\Bigr)
   \Bigl(\frac{\alpha_i}{\alpha_j}-\frac{\alpha_j}{\alpha_i}\Bigr)=0\\
\end{aligned}
\]
becomes reducible in the following cases:
\begin{align*}
 & (v_i\mp v_j)(v\mp v_{i,j})=0 &&  \text{at~~} \alpha_j=\pm\alpha_i,\\
 & (v\mp v_j)(v_i\mp v_{i,j})=0 &&  \text{at~~} \alpha_j=\pm1, \\
 & vv_iv_jv_{i,j}=0 && \text{at~~} \alpha_j=0,~\delta\ne0, \quad
 \text{after the inversion~~}
 v\to1/v.
\end{align*}

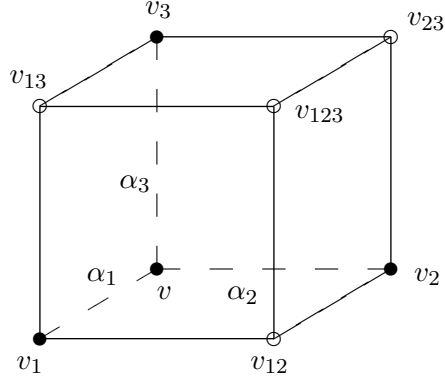
\begin{figure}[t]
\begin{center}\setlength{\unitlength}{0.08em}
\begin{picture}(200,170)(-30,-20)
 \put(100,  0){\circle{6}}  \put(0  ,100){\circle{6}}
 \put( 50, 30){\circle*{6}} \put(150,130){\circle{6}}
 \put(  0,  0){\circle*{6}} \put(100,100){\circle{6}}
 \put( 50,130){\circle*{6}} \put(150, 30){\circle*{6}}
 \path(100,0)(0,0)(0,100)(100,100)(100,0)(150,30)(150,130)(50,130)(0,100)
 \path(100,100)(150,130)
 \dashline{10}(0,0)(50,30)(50,130)
 \dashline{10}(50,30)(150,30)
 \put(-10,-13){$v_1$}
 \put(90,-13){$v_{12}$}
 \put(50,17){$v$}
 \put(-13,110){$v_{13}$}
 \put(160,25){$v_2$}
 \put(45,140){$v_3$}
 \put(109,95){$v_{123}$}
 \put(157,135){$v_{23}$}
 \put(20,25){$\alpha_1$}
 \put(80,17){$\alpha_2$}
 \put(34,65){$\alpha_3$}
\end{picture}
\caption{Three-dimensional consistency}\label{fig:cube}
\end{center}
\end{figure}

The classification in \cite{ABS} is based on the definition of integrability as
the {\em 3-dimensional consistency} condition \cite{BS,NW,Ni}. This means that
equation (\ref{Qeq}) can be consistently embedded into a 3-dimensional lattice,
so that the similar equations hold for all six faces of any elementary cube, as
on Fig.~\ref{fig:cube}. It should be mentioned that this property is in a close
relation to the {\em set-theoretical solutions of the Yang--Baxter equation}
\cite{D}, or {\em Yang--Baxter maps} \cite{V02} (see the discussion of this in
the Conclusions of \cite{ABS}).

To describe more precisely what does the 3D consistency mean, consider the
Cauchy problem with the initial data $v,v_1,v_2,v_3$. The equations on the rear
faces
\begin{equation}\label{Qij}
 Q(v,v_i,v_j,v_{ij};\alpha_i,\alpha_j)=0, \quad 1\le i<j\le 3
\end{equation}
allow one to determine uniquely the values $v_{12},v_{13},v_{23}$. After that
one has three different equations for $v_{123}$, coming from the front faces,
and consistency means that all three values thus obtained for $v_{123}$
coincide. For example, in the discrete potential KdV case we have the formula
\begin{equation}\label{v123}
 v_{123}=\frac{(\alpha_1-\alpha_2)v_1v_2
               +(\alpha_3-\alpha_1)v_3v_1+(\alpha_2-\alpha_3)v_2v_3}
              {(\alpha_3-\alpha_2)v_1
               +(\alpha_1-\alpha_3)v_2+(\alpha_2-\alpha_1)v_3}
\end{equation}
independently on the order of calculations.

It should be noted that the variable $v$ does not appear in this  expression.
This property is very important in many aspects, in particular, in the
classification problem which was solved in \cite{ABS} only under this
assumption. The role of this property for the Cauchy problems is not very
clear. Although we will not use it in our analysis of the IVP below we will see
that some results for the linear wave equation (which satisfies all the
properties (i)--(iii) but not the $v$-independence) and for the equations from
the list \cite{ABS} will be different. Of course, linear equation is
exceptional from many points of view, so it does not mean that this is related
to this particular property.

The following Theorem is the immediate consequence of 3D consistency. It
defines the B\"acklund transformation for quad-graph equation, and we will make
use of it in the next Section.

\begin{figure}[t]
\centerline{\includegraphics[width=60mm]{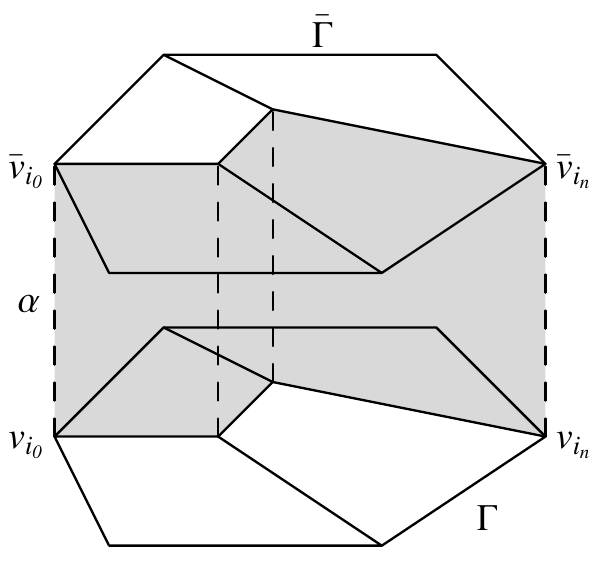}}
\caption{B\"acklund transformation}\label{fig:QeqBT}
\end{figure}

\begin{theorem}\label{th:QeqBT}
Let $v_i$, $i\in V(\Gamma)$ be a nonsingular solution of a 3D consistent
equation (\ref{Q}) from the list \cite{ABS} on a properly labelled finite
quad-graph $\Gamma$. Let us fix some vertex $i_0$ and choose an additional
parameter $\alpha$ such that all equations
\begin{equation}\label{QeqBT}
 Q(v_i,v_j,\bar v_i,\bar v_j;\alpha_{ij},\alpha)=0,\quad (i,j)\in E(\Gamma)
\end{equation}
are irreducible. Then for generic value $\bar v_{i_0}$ the relations
(\ref{QeqBT}) define uniquely a nonsingular solution $\bar v_i$ of equation
(\ref{Q}).
\end{theorem}
\begin{proof}
Let us consider some path from $i_0$ to the given vertex $i_n$. Along this
path we have the sequence of equations of the form
\[
 Q(v_{i_m},v_{i_{m+1}},\bar v_{i_m},\bar v_{i_{m+1}};\alpha_{i_m,i_{m+1}},\alpha)=0,
 \quad m=0,\dots,n-1.
\]
The solutions of these equations may turn to be singular, but one can show that
for generic $\alpha,\bar v_{i_0}$ the singularity may occur only with respect
to the variables $v_{i_k}$, while the variables $\bar v_{i_1},\dots,\bar
v_{i_n}$ can be found recursively in unique way. The fact that the different
choice of the path gives the same result and that the constructed $\bar v_i$ is
a nonsingular solution of (\ref{Q}) follows from the 3D consistency.
\end{proof}

As it was observed in \cite{BS,NW,Ni} there is a direct link between the
B\"acklund transformation (\ref{QeqBT}) and the {\em discrete zero curvature
(or Lax) representation} for the equation (\ref{Qeq}). We will call the matrix
$L(x,y;\alpha,\lambda)$ the {\em Lax matrix} for the equation (\ref{Qeq}) if
this equation is equivalent to the relation
\begin{equation}\label{ZCR}
   L(v_1,v_{12};\alpha_2,\lambda)L(v,v_1;\alpha_1,\lambda)
 = L(v_2,v_{12};\alpha_1,\lambda)L(v,v_2;\alpha_2,\lambda)
\end{equation}
for any value of the spectral parameter $\lambda$ (cf. \cite{A, BS}).

The general procedure \cite{BS,NW,Ni} allows to construct for 3D consistent
equations  $2\times2$ Lax matrices in the following way. Put $\alpha=\lambda$
and rewrite the formula (\ref{QeqBT}) in the form of the discrete Riccati
equation
\[
 \bar v_j= R(\bar v_i),\quad (i,j)\in E(\Gamma),
\]
where the function $R$ is linear fractional due to the property (iii), with the
coefficients depending on $v_i,v_j,\alpha_{ij},\lambda$. This equation can be
linearized by the substitution $\bar v=\psi/\phi$, which gives us the Lax
matrix. In the discrete KdV case this procedure brings to the following
matrices $L_{m,n}=L(v_n,v_m;\alpha_{mn},\lambda)$:
\begin{equation}\label{dpKdVL}
 L_{m,n}=\begin{pmatrix}
   -v_n & v_nv_m+\alpha_{mn}-\lambda \\
    -1  &        v_m
 \end{pmatrix}.
\end{equation}
We will use these matrices to investigate the IVP for the dpKdV equation in
section \ref{s:LL}.

\section{Initial value problems}\label{s:ivp}

Let us consider the initial value problem (or Cauchy problem) for a discrete
equation on a quad-graph $\Gamma$ with initial data on a connected subgraph
$P$. In most of the examples we consider $P$ is a {\em simple path} which is a
connected sequence of the edges of the graph without self-intersections.

\begin{definition}
A Cauchy problem for equation (\ref{Q}) on a finite, simply-connected, planar
quad-graph $\Gamma$ with a proper labelling and initial data on a connected
subgraph $P$ is called {\em well-posed} if a nonsingular solution exists and
is unique for generic initial data.
\end{definition}

It should be stressed that the singular solutions are not taken into account.
For instance, the Cauchy problem is not well-posed even if it admits a unique
solution, but it is singular for generic initial data. Conversely, if the
problem admits several singular solutions apart from the unique nonsingular
one then the problem is considered well-posed.

\begin{figure}[t]
\centerline{\includegraphics[width=80mm]{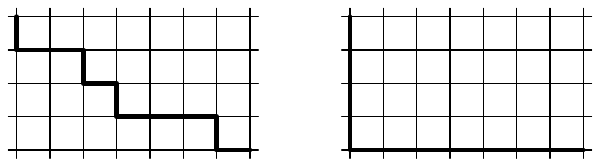}}
\captionsetup{width=0.95\textwidth}
\caption{Possible choices of initial data on the square grid}\label{fig:rect}
\end{figure}

Let us first compare the number of equations and unknowns for the given
quad-graph $\Gamma$. We assume that $\Gamma$ is simply connected and finite,
and contains $F$ faces and $V$ vertices, from which $V_b$ ones belong to the
boundary. Now apply the double counting of the angles in all faces (we assume
for simplicity that each edge is a straight line, which, of course, is not
necessary; the Euler formula provides the universal method). Since the sum of
angles in each quadrilateral is equal to $2\pi$, hence the total sum is equal
to $2\pi F$. On the other hand, summation over the vertices gives $2\pi(V-V_b)$
for the interior nodes plus $\pi(V_b-2)$ for the boundary, so that the
following equality holds:
\[
 F=V-\frac{1}{2}V_b-1.
\]
This means that in order to balance the number of equations and unknowns one
has to to assign initial data in some $\frac{1}{2}V_b+1$ vertices. These
vertices are not necessarily lie on the boundary, as the simplest example of a
staircase in rectangular domain of the square lattice demonstrates (see
Fig.~\ref{fig:rect}; here and everywhere initial data are marked by solid
line).

\paragraph{Remark.} The same calculation in the case of planar graph with
$k$-lateral faces yields the formula $(k-2)F=2V-V_b-2$. This means (under the
natural assumption that $V_b<<V$ for large domains) that the systems on the
graphs with triangular faces are overdetermined, while those on the graphs with
$k>4$ are underdetermined. Of course, the balance between equations and
unknowns can be restored in many ways. For example, one can consider systems
with variable number of equations per face. In this way some integrable models
were obtained on the triangular and hexagonal lattices, see e.g.~\cite{BHS}.
Another possibility is to consider graphs with various types of the faces.
\medskip

Of course the above consideration is too rough and cannot answer how to choose
the correct settings of Cauchy problem even in the simplest situations. In
order to get some experience let us consider several examples assuming that
equation (\ref{Qeq}) satisfies the properties (i)--(iii) from the previous
Section, but {\em is not}\/ necessarily integrable at first.

A priori, several cases are possible: the IVP may be well-posed, or it may be
underdetermined or overdetermined, and the calculation scheme for the
well-posed IVP may be explicit (need only solving of one basic equation
(\ref{Qeq}) at each step) or implicit (need to solve the systems of such
equations). It turns out that all these possibilities can be easily realized on
very simple quad-graphs.

\begin{figure}[t]
\centerline{\includegraphics[width=100mm]{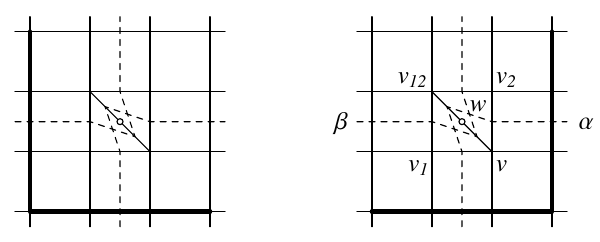}}
\caption{Well-posed and overdetermined Cauchy problems}\label{fig:defect2}
\end{figure}

\paragraph{Example 1.} It is easy to see that Cauchy problem shown on the left
Fig.~\ref{fig:defect2} is well-posed, that is all fields are uniquely defined
for generic initial data. Another choice of initial data for the same graph, as
shown on the right figure, leads to an implicit scheme and the Cauchy problem
is not well-posed. The values  $v$, $v_1$, $v_2$ are found uniquely and
unknowns $w$, $v_{12}$ satisfy the system (use symmetry (\ref{Qsym})):
\[
 Q(v,v_1,w,v_{12};\alpha,\beta)=0,\quad Q(v,v_2,w,v_{12};\alpha,\beta)=0.
\]
In general, this system possesses solutions, but only singular. Indeed, if a
solution is nonsingular then $v_1$ and $v_2$ can be found by the given
$v,w,v_{12}$, but this leads to the equality $v_1=v_2$ which is a constraint
for initial data. Therefore, there are no nonsingular solutions in the generic
case $v_1\ne v_2$. Moreover, if we consider initial data such that $v_1=v_2$,
then the solution is not unique, because only one equation for $w$ and $v_{12}$
remains. The structure of the special solution can be easily understood for a
concrete example of cross-ratio equation
\begin{equation}\label{Q01}\tag{$Q^0_1$}
 Q(v,v_i,v_j,v_{ij};\alpha_i,\alpha_j)
 =\alpha_i(v-v_j)(v_i-v_{ij})-\alpha_j(v-v_i)(v_j-v_{ij})=0.
\end{equation}
The singular solutions of this equation are characterized by coincidence of, at
least, three variables. Indeed, the above system at $v_1\ne v_2$ is solved by
$w=v_{12}=v$.

\begin{figure}[t]
\centerline{\includegraphics[width=100mm]{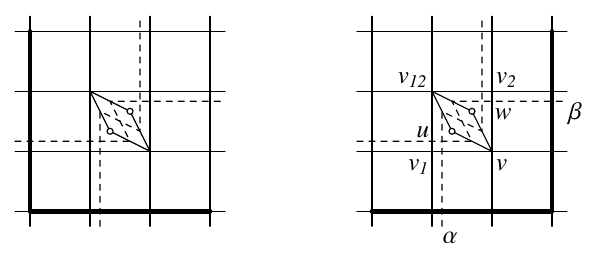}}
\captionsetup{width=0.8\textwidth}
\caption{Well-posed Cauchy problems with explicit and implicit
calculation scheme on transparent defect}\label{fig:defect22}
\end{figure}

\paragraph{Example 2.} Both IVP shown on the Fig.~\ref{fig:defect22} are
well-posed. For the left one this is obvious; for the right one we have the
system of three equations on the variables $u,w,v_{12}$:
\[
 Q(v,v_1,u,v_{12};\alpha,\beta)=0,\quad
 Q(v,w,u  ,v_{12};\alpha,\beta)=0,\quad
 Q(v,w,v_2,v_{12};\alpha,\beta)=0.
\]
This system implies $u=v_2$, $w=v_1$ and for $v_{12}$ one has the same equation
$Q(v,v_1,v_2,v_{12};\alpha,\beta)=0$ as in the case of lattice without defect.
In other words, the defect shown on these pictures does not affect the
dynamics. We shall call defects with such property {\em transparent}.

Analogously, the graph shown on the Fig.~\ref{fig:defect_S} gives explicit and
implicit schemes depending on the choice of initial data.

\def\QG{\multiput(0,0)(0,20){4}{\path(-5,0)(105,0)}
\path(0,-5)(0,65)   \path(20,-5)(20,65)
\path(80,-5)(80,65) \path(100,-5)(100,65)
\path(40,-5)(40,20) \path(40,33)(40,65)
\path(60,-5)(60,27) \path(60,40)(60,65)
\path(20,40)(30,33)(40,20)
\path(30,33)(40,33)(60,20)
\path(40,40)(60,27)(70,27)
\path(60,40)(70,27)(80,20)
\dashline{3}(30,-5)(30,20)(25,36.5)(40,36.5)(60,23.5)(75,23.5)(70,40)(70,65)
\dashline{3}(-5,26.5)(50,26.5)(50,33.5)(105,33.5)
}
\begin{figure}[t]
\begin{center}\setlength{\unitlength}{0.12em}
\begin{picture}(130,90)(-15,-10)
 \QG \bfput{\path(0,60)(0,0)(100,0)}\put(20,-15){\em a}
\end{picture}
\begin{picture}(130,90)(-15,-10)
 \QG \bfput{\path(0,0)(100,0)(100,60)}\put(20,-15){\em b}
\end{picture}
\caption{Well-posed IVP with explicit (a) and implicit (b) schemes}
\label{fig:defect_S}
\end{center}
\end{figure}
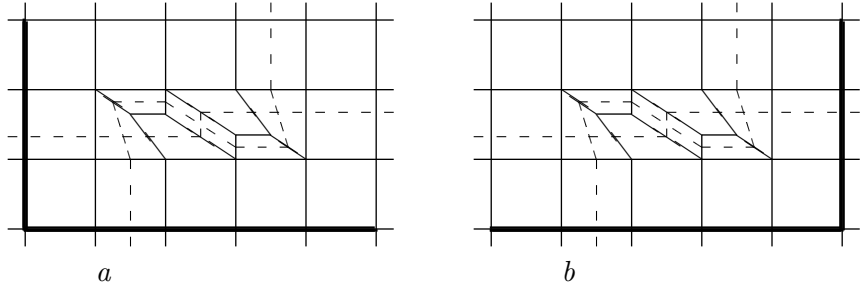

\begin{figure}[t]
\centerline{\includegraphics[width=110mm]{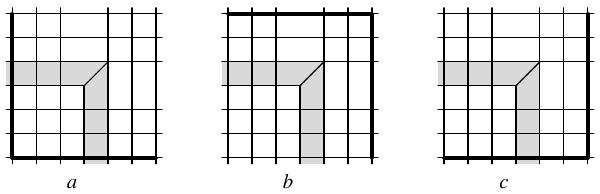}}
\captionsetup{width=0.95\textwidth}%
\caption{Overdetermined, underdetermined and well-posed Cauchy problems}
\label{fig:defect3}
\end{figure}

\paragraph{Example 3.} For the quad-graph shown on the Fig.~\ref{fig:defect3}
the first two IVP are not well-posed: for a) case one finds successfully  the
values $v$ inside the square bounded by initial data and the dashed
characteristic, but encounters contradiction outside it; for b) case the same
square remains undetermined.

\def\QG{
\multiput(0, 0)(20,-10){5}{\path(-6,-3)(80,40)(80,105)}
\multiput(0, 0)(20, 10){4}{\path(0,65)(0,0)(86,-43)}
\multiput(0, 0)( 0, 20){4}{\path(-6,-3)(80,40)(166,-3)}
\dashline{3}(-6,7)(80,50)(166,7)
\dashline{3}(4,-8)(90,35)(90,101)
\dashline{3}(70,101)(70,35)(154,-8)}

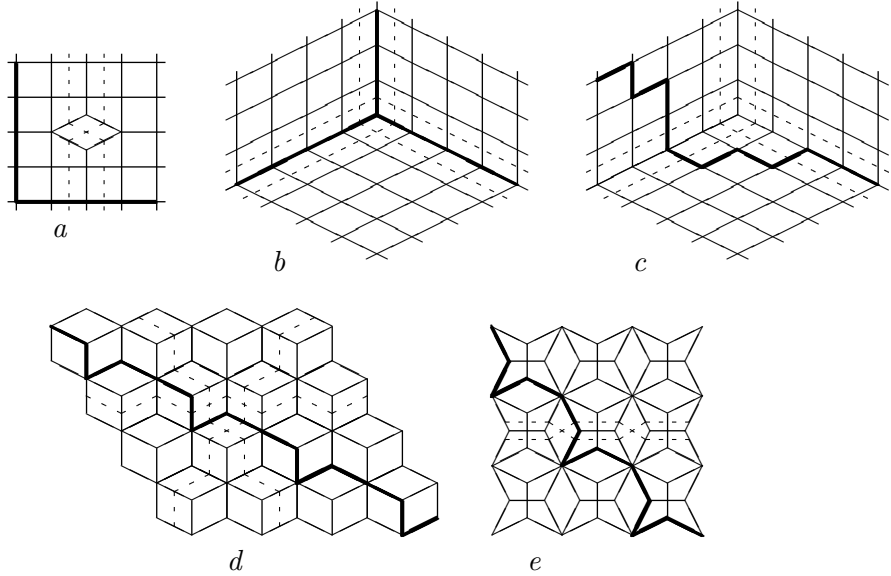
\begin{figure}[t]
\begin{center}\setlength{\unitlength}{0.06em}
\begin{picture}(120,160)(-30,-30)
 \multiput(0, 0)(0,20){2}{\path(-5,0)(85,0)}
 \multiput(0,60)(0,20){2}{\path(-5,0)(85,0)}
 \multiput(0, 0)(20,0){2}{\path(0,-5)(0,85)}
 \multiput(60,0)(20,0){2}{\path(0,-5)(0,85)}
 \path(-5,40)(20,40)(40,30)(60,40)(40,50)(20,40)
 \path(40,-5)(40,30)\path(40,50)(40,85)\path(60,40)(85,40)
 \dashline{3}(30,85)(30,45)(50,35)(50,-5)
 \dashline{3}(50,85)(50,45)(30,35)(30,-5)
 \bfput{\path(0,80)(0,0)(80,0)}
 \put(20,-20){\em a}
\end{picture}
\begin{picture}(200,160)(-30,-40)
 \QG \bfput{\path(0,0)(80,40)(160,0)\path(80,40)(80,100)}\put(20,-50){\em b}
\end{picture}
\begin{picture}(200,160)(-30,-40)
 \QG \bfput{\path(0,60)(20,70)(20,50)(40,60)(40,20)(60,10)
 (80,20)(100,10)(120,20)(160,0)}
 \put(20,-50){\em c}
\end{picture}
\def\QG{\path( 0,0)(-20,10)(-20,30)(0,20)(0,0)(20,10)(20,30)(0,20)
\path(-20,30)(0,40)(20,30)}
\begin{picture}(200,170)(-70,-10)
 \multiput(0,0)(-20,30){4}{\multiput(0,0)(40,0){4}{\QG}}
 \multiput(-60,60)(40,0){4}{\dashline{3}(0,20)(20,10)(40,20)}
 \multiput(-10,5)(20,30){4}{\dashline{3}(0,0)(0,20)(20,30)}
 \multiput(50,5)(-20,30){4}{\dashline{3}(0,0)(0,20)(-20,30)}
 \bfput{\multiput(-80,120)(60,-30){3}
 {\path(0,0)(20,-10)(20,-30)(40,-20)(60,-30)}
 \path(100,30)(120,20)(120,0)(140,10)}
 \put(20,-20){\em d}
\end{picture}
\def\QG{\path( 0,0)(20,10)(40, 0)
\path(0,40)(20,30)(40,40)
\path( 0,0)(10,20)( 0,40)
\path(40,0)(30,20)(40,40)
\path(10,20)(30,20)\path(20,10)(20,30)}
\begin{picture}(200,170)(-30,-10)
 \multiput(0,0)(40,0){3}{\multiput(0,0)(0,40){3}{\QG}}
 \multiput(7.5,65)(40,0){3}{\dashline{3}(0,0)(25,0)\dashline{3}(0,-10)(25,-10)}
 \multiput(32.5,65)(40,0){2}{\dashline{3}(0,0)(15,-10)\dashline{3}(0,-10)(15,0)}
 \bfput{\multiput(0,80)(40,-40){3}{\path(0,40)(10,20)(0,0)(20,10)(40,0)}}
 \put(20,-20){\em e}
\end{picture}
\caption{Examples of well-posed Cauchy problems}\label{fig:qg}
\end{center}
\end{figure}

\paragraph{Example 4.} Several examples of well-posed IVP on different
quad-graphs are shown on the Fig.~\ref{fig:qg}.

Inspired by these examples and analogies with continuous case one could
conjecture that an IVP with initial data given on a simple path $P$ is
well-posed iff each characteristic (or strip) in $\Gamma$ meets $P$ exactly at
one edge. We finish this section with two examples  which show that for the
general equation this conjecture is not true and emphasize the role of the 3D
consistency condition in this problem.

\begin{figure}[t]
\centerline{\includegraphics[width=70mm]{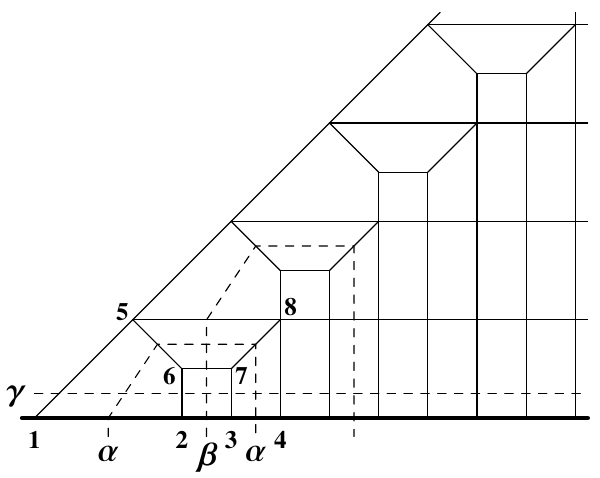}}
\captionsetup{width=0.75\textwidth}
\caption{Well-posed/overdetermined Cauchy problem in general/integrable case.}
\label{fig:defect10}
\end{figure}

\paragraph{Example 5.} On the quad-graph on the Fig.~\ref{fig:defect10}, the
vertical characteristics come back and cross the initial data again, while the
horizontal characteristics do not cross the initial data at all. Nevertheless,
this IVP is well-posed in general case, by means of implicit  calculation
scheme which can be described as follows:

1) the implicit step: unknowns $v_5$, $v_6$, $v_7$, $v_8$ are found as
functions on $v_1$, $v_2$, $v_3$, $v_4$ by solving the system
\begin{equation}\label{sys8}
\begin{gathered}
 Q(v_1,v_2,v_5,v_6;\alpha,\gamma)=0,\quad
 Q(v_2,v_3,v_6,v_7;\beta,\gamma)=0,\\
 Q(v_3,v_4,v_7,v_8;\alpha,\gamma)=0,\quad
 Q(v_6,v_7,v_5,v_8;\beta,\alpha)=0.
\end{gathered}
\end{equation}

2) next, all other variables on the bottom strip are computed explicitly;

3) now we are in the same situation and can repeat the procedure.

Of course, there is a question whether above system is degenerate or not. There
is no reason for it to be degenerate for a generic affine-linear polynomial
$Q$. It is clear that solving first three equations allows to express the
unknowns $v_6,v_7,v_8$ as rational functions of the other ones and then the
rest equation turns into an equation of degree 4 with respect to $v_5$.
Therefore, this Cauchy problem can be considered well-posed for a generic
equation, with a stipulation that we allow multivaluedness on each step of
implicit scheme.

The situation is different if the equation is 3D-consistent. Let us consider
again equation (\ref{Q01}) as an example. It turns out that if the initial data
$v_1,v_2,v_3,v_4$ are generic then system (\ref{sys8}) possesses two solutions
of multiplicity 2, but both solutions are singular. One of them is
\[
 v_5=v_6=v_7=v_2,\quad
 v_8=\frac{\gamma v_2(v_3-v_4)+\alpha v_4(v_2-v_3)}
  {\gamma(v_3-v_4)+\alpha(v_2-v_3)},
\]
and the other one is its mirror image with respect to the strip $\beta$.

The origin of this degeneration can be easily understood. Let us identify the
faces involved in the system (\ref{sys8}) with four faces of a cube. Assume
that this system admits a nonsingular solution, then all unknowns can be
expressed through, say, $v_6$, $v_2$, $v_5$ and $v_7$. The 3D-consistency
implies that analogous equation are fulfilled on the rest two faces, that is
$Q(v_1,v_2,v_4,v_3;\alpha,\beta)=0$ and $Q(v_1,v_4,v_5,v_8;\beta,\gamma)=0$.
But, first of these equations is a constraint for initial data, therefore the
system cannot posses nonsingular solutions for the generic initial data.
Moreover, if we impose this constraint then the solution becomes nonsingular,
but also not unique (compare with example 1). Indeed, in such a case one of
unknowns $v_5,v_6,v_7,v_8$ can be chosen arbitrarily, the other ones are found
from three equations of the system and the fourth equation will be satisfied
automatically due to the 3D-consistency. Thus, in this example the Cauchy
problem for 3D-consistent equation is not well-posed.

\begin{figure}[t]
\centerline{\includegraphics[width=40mm]{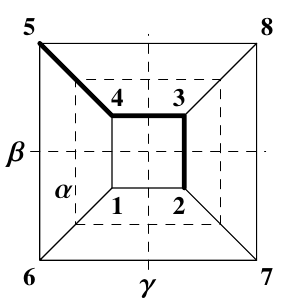}}
\captionsetup{width=0.75\textwidth}
\caption{Overdetermined/well-posed Cauchy problem in general/integrable case.}
\label{fig:aaa}
\end{figure}

\paragraph{Example 6.} The finite quad-graph on the Fig.~\ref{fig:aaa} gives an
example of the opposite situation: the Cauchy problem is well-posed for
integrable equation and overdetermined for the generic one. In this example we
have three characteristics, each crossing the path with initial data
$v_2,v_3,v_4,v_5$ at one edge. The value $v_8$ can be computed in two ways:
directly from the data $v_3,v_4,v_5$ or in several steps after finding
consequently $v_1,v_6$ and $v_7$. In general, this values are different, so
that the problem is overdetermined. However, for a 3D consistent equation both
values coincide: the Fig.~\ref{fig:aaa} is nothing but the projection of the
cube on the Fig.~\ref{fig:cube}.

\section{Existence and uniqueness theorem}\label{s:corr}

Now we are ready to formulate our main result.

\begin{theorem}\label{th:corr}
Let equation (\ref{Qeq}) be 3D consistent, and let $P$ be a simple path in a
finite simply connected planar quad-graph $\Gamma$ with a proper labelling.
Consider Cauchy problem for this equation with generic initial data on
the path $P$.

1) If each characteristic in $\Gamma$ intersects $P$ exactly at one edge then
the IVP is well-posed, that is a nonsingular solution exists and is unique.

2) If some characteristic intersects $P$ more than once then the IVP is
overdetermined (no nonsingular solution exists for generic initial data).

3) If some characteristic does not intersect $P$ then the IVP is
underdetermined (if a nonsingular solution exists then it is not unique).
\end{theorem}

We should clarify that when we consider the intersections of  characteristics
with a path we count only the internal edges of the characteristics. For
example the intersection of a vertical characteristic with a staircase is one
edge (but not two).

\begin{Proof}[existence]
Assume that the conditions of the part 1) of the Theorem are fulfilled. Let us
construct an immersion of $\Gamma$ into the $N$-dimensional unit cube, where
$N$ is the length of $P$. To do this, assign the shift operators $T_i$,
$i=1,\dots,N$ to the edges of $P$. By definition, $T_i$ acts on $N$-dimensional
vectors by formula $T_i(x)=x+e_i \pmod2$ where $e_i$ is the vector with 1 at
$i$-th place and 0 at the others.

Now distribute these operators at all edges of $\Gamma$ according to the same
rule as for the parameters $\alpha$, i.e.~one operator per strip. Obviously
this can be done in unique way due to the condition 1).  Choose the first
vertex $v_0$ of $P$ as the origin with coordinates $(0,\dots,0)$ and define the
coordinates of any other vertex $v$ as the result of applying shift operators
along any path $P_v$ from $v_0$ to $v$.  The result does not depend on the
choice of $P_v$. Indeed, any closed path has an even intersection index with
any characteristic and therefore is equivalent to identity operator.

Thus we have found an immersion of $\Gamma$ into $N$-dimensional unit cube and
each face of $\Gamma$ now is just a 2-dimensional face in this cube. Initial
values are given on the path
\[
 P=((0,0,\dots,0),\;(1,0,\dots,0),\;(1,1,\dots,0),~\dots,\;(1,1,\dots,1))
\]
and allow to calculate the values in each vertex of the cube. (It is not
difficult to explain how to obtain the values for the coordinate vectors $e_i$,
and so on.) Of course, 3D consistency is necessary here in order to justify this procedure. 
In this way we construct some nonsingular
solution of the Cauchy problem.
\end{Proof}

\paragraph{Remark.} It is worthy to note that all the fields at the vertices of
the multidimensional cube can be calculated recursively using only one equation
each time, so that on the cube the integration scheme becomes {\em explicit}.
This means, in particular, that the solution in each vertex of quad-graph is a
rational function of initial data. This algorithm of solving Cauchy problem is
not very effective, since the number $2^N$ of the vertices in the
$N$-dimensional cube grows exponentially in $N$, while the number of vertices
in $\Gamma$ (and therefore the size of the algebraic system we are solving) is
about $N^2$. From the practical point of view, one should look for an embedding
of quad-graph into a lattice $\Integer^d$ of minimal dimension.
\bigskip

The proof of the rest of the theorem presented below is based on the very
different ideas and uses the notion of B\"acklund transformation (see Theorem
\ref{th:QeqBT}). We will transform the initial quad-graph by erasing and
inserting of a strip. These transformations bring, roughly speaking, to an
equivalent IVP and can be described as follows.

\begin{figure}[t]
\centerline{\includegraphics[width=100mm]{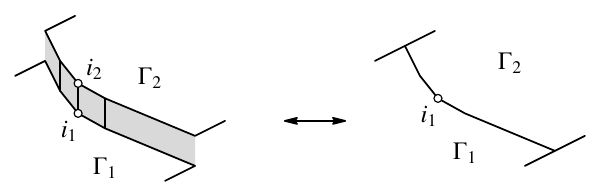}}
\caption{Erasing and inserting of a strip}\label{fig:cutting}
\end{figure}

Let us consider some strip $C$ with parameter $\alpha$. It divides $\Gamma$
into two subgraphs, $\Gamma_1$ and $\Gamma_2$. Let us denote $P_i$ the
corresponding boundaries of the strip and let $i_1\in P_1$ and $i_2\in P_2$ be a
pair of neighbour vertices separated by $C$. Now, let us apply BT (\ref{QeqBT})
to the part $\Gamma_2$ of the quad-graph, taking $i_2$ as the seed point, and
$\bar v_{i_2}=v_{i_1}$. This will give us some new solution in $\Gamma_2$ such
that the variables $v$ along the boundary $P_2$ coincide  with the corresponding
variables $v$ along $P_1$. Therefore we can construct the common solution on
the graph $\widetilde\Gamma$ obtained from $\Gamma$ by removing of the strip
$C$. This solution coincides with the old one on $\Gamma_1$. Of course, some
information is lost after such an operation and in order to make it invertible
one have to remember the value $v_{i_2}$.

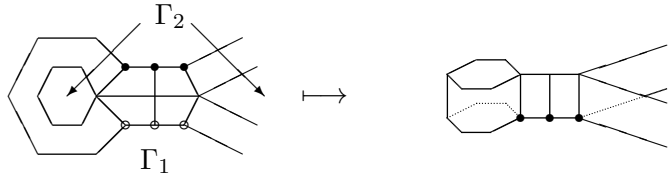
\begin{figure}[b]
\begin{center}
\setlength{\unitlength}{0.05em}
\begin{picture}(450,140)(0,-70)
 \path(160,-40)(120,-20)(80,-20)(60,-40)(20,-40)(0,0)(20,40)(60,40)(80,20)(120,20)(160,40)
 \path(60,0)(50,-20)(30,-20)(20,0)(30,20)(50,20)(60,0)(130,0)(170,20)
 \path(130,0)(170,-20)
 \path(80,-20)(60,0)(80,20) \multiput(80,20)(20,0){3}{\circle*{5}}
 \path(100,-20)(100,20)     \multiput(80,-20)(20,0){3}{\circle{5}}
 \path(120,-20)(130,0)(120,20)
 \put(90,-50){$\Gamma_1$} \put(100,50){$\Gamma_2$}
 \put(90,50){\vector(-1,-1){50}} \put(125,50){\vector(1,-1){50}}
 \put(200,-5){$\longmapsto$}
 \put(300,-15){\path(0,0)(0,30)(20,40)(40,40)(50,30)(50,0)
  \path(70,0)(70,30)\path(90,0)(90,30)(150,50)
  \path(150,20)(135,15)\dottedline{2}(135,15)(90,0)
  \multiput(0,0)(0,30){2}{\path(0,0)(10,-10)(30,-10)(50,0)(90,0)(150,-20)}
  \dottedline{2}(0,0)(20,10)(40,10)(50,0)
  \multiput(50,0)(20,0){3}{\circle*{5}}}
 \end{picture}
\caption{Case of self-tangent strip}\label{fig:selftangent}
\end{center}
\end{figure}

On the Fig.~\ref{fig:cutting} these operations are shown for the case of simple
strip, however the closed or self-tangent strips are allowed as well. For the
self-tangent strips the 3D consistency property implies the coincidence of the
fields in some vertices, and these vertices can be merged during the above
procedure (Fig.~\ref{fig:selftangent}).

\begin{Proof}[necessity]{\itshape\bfseries 2)}
Let some characteristic $C$ intersect the path $P$ with initial data twice.
Let us enumerate the vertices of $P$ as shown on the Fig.~\ref{fig:nec1}, so
that the edges of intersection are $(1,2)$ and $(n-1,n)$.

\def\QG{\path(0,0)(-10,10)(0,30)\path(70,0)(90,10)(80,20)
 \multiput(30,20)(10,0){3}{\circle*{2}}
 \put(10,-5){\circle*{2}}
 \put(20,-10){\circle*{2}}
 \put(30,-12){\circle*{2}}
 \put(40,-12){\circle*{2}}
 \put(50,-10){\circle*{2}}
 \put(60,-5){\circle*{2}}
 {\put(30,0){$\Gamma_1$}\scriptsize
 \put(3,31){$2$}\put(70,13){$n-1$}}}
\begin{figure}[htb]
\begin{center}\setlength{\unitlength}{0.10em}
\begin{picture}(180,70)(-20,-20)
 \QG
 \path(0,0)(-20,0)(-30,10)(-10,10)
 \path(-30,10)(-20,30)(0,30)
 \path(70,0)(100,0)(110,10)(90,10)
 \path(110,10)(100,30)(80,20)
 \dashline{3}(-8,32)(-20,10)(-8,-2)
 \dashline{3}(88,27)(100,10)(88,-2)
 {\scriptsize
 \put(-20,33){$1$}\put(100,32){$n$}
 \put(-10,-8){$\alpha$}\put(80,-8){$\alpha$}}
 \bfput{\path(-20,30)(0,30)(20,20)\path(60,20)(80,20)(100,30)}
 \put(130,15){$\longmapsto$}
\end{picture}
\begin{picture}(100,70)(-20,-20)
 \QG
 \path(0,0)(0,20)(-10,30)(0,50)(20,40)(20,20)
 \path(-10,10)(-10,30)
 \path(70,0)(70,20)(90,30)(80,40)(60,40)(60,20)
 \path(90,10)(90,30)
 \dashline{3}(2,8)(-10,20)(0,40)(20,30)(25,30)
 \dashline{3}(68,9)(90,20)(80,30)(55,30)
 {\scriptsize \put(-5,51){$1$}\put(82,41){$n$}\put(40,30){$\alpha$}}
 \bfput{\path(0,50)(0,30)(20,20)\path(60,20)(80,20)(80,40)}
\end{picture}
\caption{Constraint on the initial data}\label{fig:nec1}
\end{center}
\end{figure}
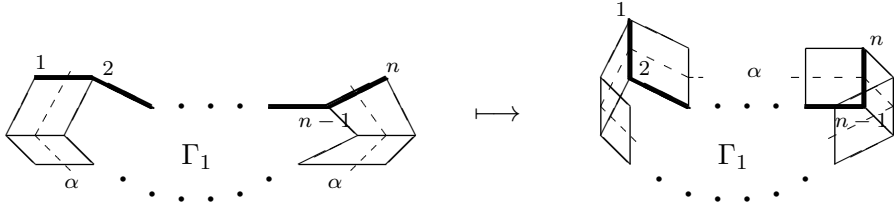

Suppose that IVP is well-posed, then a nonsingular solution of equation
(\ref{Qeq}) exists on the subgraph $\Gamma_1$ bounded by $C$ and $P$. Let us
look at this picture as 3-dimensional one, assuming that the exterior boundary of
the strip $C$ lies at the next floor, in particular the vertex $1$ is over $2$
and $n$ is over $n-1$. According to Theorem \ref{th:QeqBT} the value
$v_n$ can be defined by solving the sequence of equation (\ref{QeqBT}) along
the path $(2,\dots,n-1)$, that is $v_n$ is some function on
$v_1,\dots,v_{n-1}$. This means that the initial data are not free, that is the
IVP is overdetermined.
\medskip

{\itshape\bfseries 3)} Suppose that some strip $C$ never intersect the path $P$
with initial data. Let us cut off this strip, as shown on the
Fig.~\ref{fig:cutting} (we assume for defineteness that $P$ lies in the
subgraph $\Gamma_1$). If a nonsingular solution of the IVP exists in $\Gamma$
then this procedure gives us a nonsingular solution with the same initial data
on the reduced graph $\widetilde\Gamma$.

Next let us apply the inverse procedure to this solution, namely consider BT on
$\Gamma_2$, taking some arbitrary value for $\bar v_{i_1}$. This gives us some
solution on the graph $\Gamma$ which coincides with the original solution on
$\Gamma_1$ but not on $\Gamma_2$. This means that the initial data are not
sufficient to provide uniqueness of the nonsingular solution.
\end{Proof}

\begin{Proof}[uniqueness]
This can be obtained by sequential erasing of the strips, as explained above,
until the quad-graph is exhausted. More precisely, let us consider the
operation of contracting of the internal edges of one strip in the
quad-graph. This operation preserves the proper labelling, does not generate
new strips and does not affects on the intersection of other strips with $P$.
The other strips may only shorten by several cells, in dependence on the number
of intersections with the contracted strip (in particular, they may degenerate
into a single edge). Let us consider the strip $C$ which passes through the
first edge $(1,2)$ of the path $P$. It divides $\Gamma$ into the subgraph
$\Gamma_1$ which contains $P$ and $\Gamma_2$ which does not contain $P$. Let us
apply the B\"acklund transformation from theorem \ref{th:QeqBT} to the
restriction of solution on $\Gamma_2$, with the parameter given by the labelling
of the strip and initial value $v_2$. We glue it with the restriction of the
solution onto $\Gamma_1$ and obtain a nonsingular (for generic initial values)
solution of the Cauchy problem on the quad-graph $\Gamma'$ obtained from
$\Gamma$ by contracting the strip $C$ and with a shortened path $P'$. If there
were two different nonsingular solutions on $\Gamma$ with the same
initial data then this operation brings to different nonsingular solutions on
$\Gamma'$. Repeating the process we come to a contradiction in a finite number
of steps.
\end{Proof}

\paragraph{Remark.}  Although we have assumed that $P$ is a simple path (which
is probably the most natural case), it follows from the proof that the result
is true for any connected subgraph.

We will call a connected subgraph $P$ {\em Cauchy subgraph} if it has the
property that each characteristic in $\Gamma$ meets $P$ exactly at one edge.

A natural question is whether such a subgraph does always exist or not. The
answer in general is negative. The following picture gives of an example of a
regular lattice on which no well-posed IVP is possible, at least with initial
data on the connected subgraphs. It is obtained from the kagome lattice by the
standard process of merging with the dual graph (which is the quad-graph shown
on the Fig.~\ref{fig:qg}d). For the equations on quad-graphs of such type the
question arise how to describe the set of general solutions.

\begin{figure}[htb]
\centerline{\includegraphics[width=70mm]{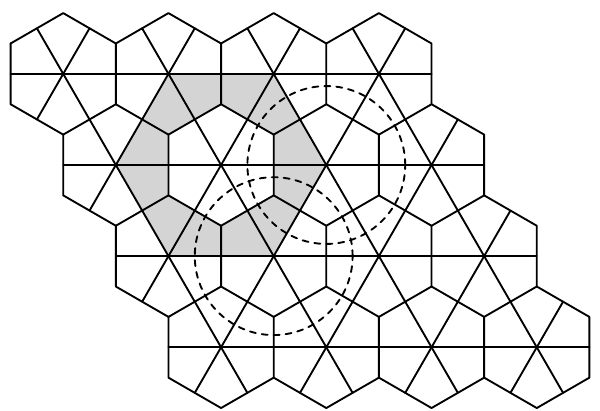}}
\caption{Example of the periodic quad-graph without connected Cauchy subgraphs:
all characteristics are closed}\label{fig:kagome}
\end{figure}

We would like to mention that as the example 6 above shows for a generic
equation the Cauchy problem with the initial data on a Cauchy subgraph could
not be well-posed. However in that example one of the characteristics is
closed. We do not know if such an example exists without closed
characteristics.

\section{Interaction of the solutions with the localized defects}\label{s:LL}

Consider the dpKdV equation
\begin{equation}\label{homdpKdVmn}
 (v_{m+1,n+1}-v_{m,n})(v_{m+1,n}-v_{m,n+1})=\alpha_m-\beta_n,\quad
 m,n\in\Integer
\end{equation}
on the regular square lattice, assuming that all strips carry different
parameters and initial data are chosen on the coordinate axes.

Let us replace some $M\times N$ rectangle inside first quadrant by a finite
quad-graph $D$ (with the same vertices on the boundary). What we will get is a
{\em regular square lattice with localized defect}. Many examples of such
localized defects can be found in the previous section, see e.g. Figures
\ref{fig:defect2}, \ref{fig:defect22}, \ref{fig:defect_S}, \ref{fig:qg}a.
Outside of the defect we have a regular picture of horizontal and vertical
strips.

According to theorem \ref{th:corr} the new Cauchy problem remains well-posed if
any strip passing through $D$ has exactly one intersection with the coordinate
semiaxes and $D$ does not contain closed strips. In particular, we will say
that the defect is {\em weak} if all the strips entering the defect leave it in
the same direction but possibly in a different order. One can see the examples
of weak defects on the Figures \ref{fig:defect22}, \ref{fig:defect_S},
\ref{fig:qg}a. On Fig.~\ref{fig:defect2}, the defect is not weak, since a pair
of characteristics change their types.

\begin{figure}[t]
\centerline{\includegraphics[width=85mm]{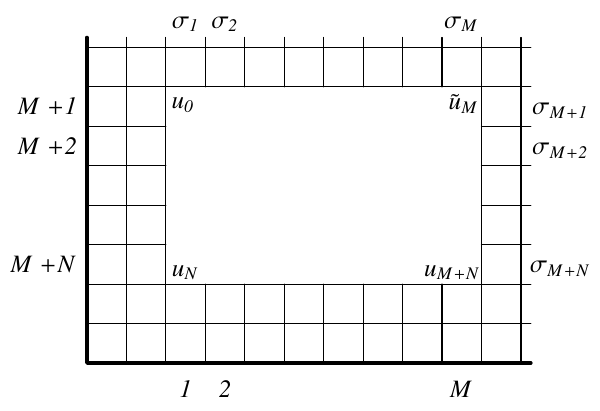}}
\caption{Permutation of strips on the boundary of the defect}\label{fig:G_defect}
\end{figure}

Now, let us compare the solution of the Cauchy problem for such localized
defect with the solution on the regular square lattice with the same initial
data. It turns out that if we are interested only in the solution outside the
defect then only the permutation of strips caused by the defect is important
while its internal structure does not matter. The permutation is defined as
follows. Consider the strips which enter into $D$ from two sides of the
rectangle as shown on Fig. \ref{fig:G_defect}) and enumerate them through
$1,\dots,M+N$. The permutation is defined then as the sequence
$(\sigma_1,\sigma_2,\dots,\sigma_{M+N})$ of the strips which leave the
rectangle from two opposite sides. On the regular square lattice these strips
form the identical permutation $(1,2,\dots,M+N)$. We will call all defects with
such permutation {\em transparent}.

\begin{figure}[t]
\centerline{\includegraphics[width=85mm]{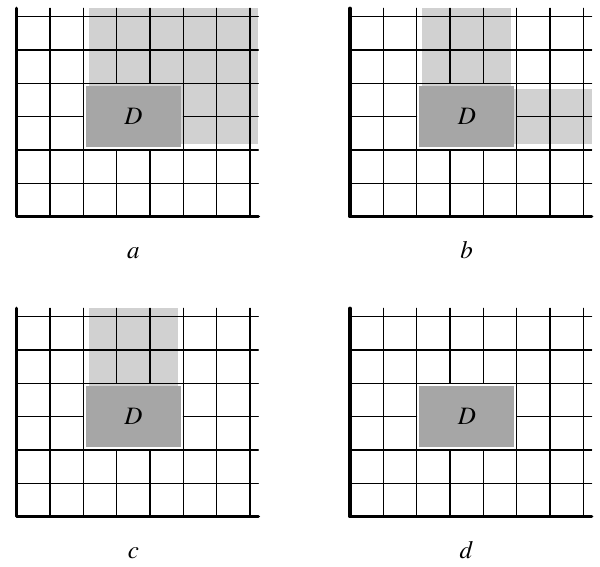}}
\captionsetup{width=0.95\textwidth}
\caption{Domains of impact on the solution of the Cauchy problem:
a) a general localized defect; b) permutation acts on vertical and horizontal
strips separately; c) permutation act on vertical strips only; d)
transparent defect}\label{fig:G_shade}
\end{figure}

\begin{theorem}\label{th:defect}
If two localized defects with the same boundary define the same
permutation of strips then the solutions of the corresponding Cauchy problems
coincide outside the defects.
\end{theorem}

In particular, a transparent defect does not affect the solution of the Cauchy
problem on the regular lattice. If the permutation defined by the defect
involves only strips of one type, say, vertical, then the solution is affected
only along these strips. Analogously, if permutation acts on vertical and
horizontal strips separately, that is $\sigma_i\le M$, $i=1,\dots,M$, then the
solution is not affected outside the union of strips passing through the defect
(see Fig. \ref{fig:G_shade}).

The proof is based on the existence of discrete zero-curvature representation
(\ref{ZCR}), (\ref{dpKdVL}) and follows from the following general
considerations.

Let $P=(i_1,i_2,\dots,i_n,i_1)$ be a closed path in a simply connected
quad-graph $\Gamma$ then for any solution of dpKdV equation on $\Gamma$ the
ordered product of the Lax matrices (\ref{dpKdVL}) is equal to a scalar matrix.
More precisely the formula
\[
 \prod^\curvearrowleft_PL_{i_{k+1},i_k}
 =\prod_C(\lambda-\alpha_C)^{\frac{1}{2}\ind(C,P)}I
\]
is valid where the product in the right hand side is taken over all
characteristics in $\Gamma$, $\alpha_C$ is the parameter corresponding to the
strip $C$ and the {\em index} $\ind(C,P)$ is the total number of intersections
of $C$ and $P$ (without signs). This formula can be easily proved from the
property (\ref{ZCR}) and the relation
\[
 L_{i,j}L_{j,i}=(\lambda-\alpha_{i,j})I.
\]
Equivalently, if $P_1=(i_1,i_2,\dots,i_m)$ and $P_2=(j_1,j_2,\dots,j_n)$ are
two paths from $i_1=j_1$ to $i_m=j_n$ then
\[
 \prod^\curvearrowleft_{P_1}L_{i_{k+1},i_k}
  =\prod_C(\lambda-\alpha_C)^{\frac{1}{2}(\ind(C,P_1)-\ind(C,P_2))}
   \prod^\curvearrowleft_{P_2}L_{j_{k+1},j_k}.
\]

\begin{theorem}\label{th:LL}
Let the paths $P_1$ and $P_2$ have the common starting and end points and for
any characteristic $C$ in $\Gamma$ we have $\ind(C,P_1)=\ind(C,P_2)\leq1$. Then
for any generic solution of dpKdV equation on $\Gamma$ the fields $v$ on $P_2$
can be uniquely recovered from the values on $P_1$.
\end{theorem}
\begin{proof}
Let us consider the matrix product
$L(\lambda)=\prod\limits^\curvearrowleft_{P_1}L_{i_{k+1},i_k}
=\left(\begin{smallmatrix}a&b\\c&d\end{smallmatrix}\right)$.  One can prove
that for a generic solution $\mathop{\rm rank}L(\alpha_C)=1$ for each strip $C$
intersecting $P_1$ (for some special solutions this can fail: for example if we
have $\alpha_{i_{k+2},i_{k+1}}=\alpha_{i_{k+1},i_k}=\alpha$ and
$v_{i_{k+2}}=v_{i_{k+1}}$ for three consequent vertices in the path, then
$L_{i_{k+2},i_{k+1}}L_{i_{k+1},i_k}|_{\lambda=\alpha}\equiv0$).

Now the result follows from the uniqueness of the refactorization of the same
matrix as $L(\lambda)=\prod\limits^\curvearrowleft_{P_2}L_{j_{k+1},j_k}$ along
the path $P_2$. It is enough to show that there exists a unique matrix of the
form
\[
 L_{j_2,j_1}=\begin{pmatrix}
  -v_{j_1} & v_{j_1}v_{j_2}+\alpha_{j_2,j_1}-\lambda\\
     -1    & v_{j_2}
 \end{pmatrix}
\]
such that $L(\lambda)=\tilde LL_{j_2,j_1}$ and the matrix $\tilde L$ is
polynomial in $\lambda$. Indeed, $v_{j_2}$ is uniquely defined from the
condition $\ker L(\alpha_{j_2,j_1})\sim \binom{v_{i_2}}{1}$. Since the
polynomials $av_{i_2}+b$ and $cv_{j_2}+d$ are divisible by
$\lambda-\alpha_{j_2,j_1}$, hence the matrix
\[
 \tilde L=\frac{1}{\lambda-\alpha_{j_2,j_1}} \begin{pmatrix}
   -av_{j_2}-b & (\alpha_{j_2,j_1}-\lambda)a +v_{j_1}(av_{j_2}+b)\\
   -cv_{j_2}-d & (\alpha_{j_2,j_1}-\lambda)c +v_{j_1}(cv_{j_2}+d)
  \end{pmatrix}
\]
is polynomial. Continuing in the same way we will reconstruct the solution on
the path $P_2$.
\end{proof}

Now to derive the Theorem \ref{th:defect} one should consider two sides of the
rectangle $D$ directed towards the Cauchy subgraph as $P_1$ and other two sides
as $P_2$.

Notice that in general the Cauchy problem with the initial data on $P_1$ may be
underdetermined,  since some characteristics may not intersect it. For example,
some closed strips may occur inside the subgraph bounded by $P_1$ and $P_2$.
Nevertheless, the use of transition matrices $L_{ji}$ will allow us to skip
over such regions of underdeterminancy.

For a well-posed IVP the proof of Theorem \ref{th:LL} gives an integration
scheme which allows to obtain the solution of dpKdV equation by refactorization
of the $L$ matrices product along the path of initial data. It is probably the
most effective scheme one can suggest for this problem.

\section{Kinks and solitons on quad-graphs}\label{s:kink}

In this section we consider some explicit solutions of dpKdV equation rather
than solutions of Cauchy problem. The hint is to search for solution compatible
with the continuous  dynamics. The dpKdV equation on $\Integer^2$ lattice
(\ref{dpKdVmn}) originates from the nonlinear  superposition principle for the
B\"acklund transformation
\begin{equation}\label{KdVBT}
 \bar v_x+v_x=(\bar v-v)^2-\lambda
\end{equation}
of the potential KdV equation $v_t=v_{xxx}-6v^2_x$. Usage of equation
(\ref{KdVBT}) allows to obtain the profile of the kink solution, and then to
find algebraically the dependence of the phase on $m$ and $n$ (the dependence
on $t$ can be easily recovered, but now this is unessential).

\begin{figure}[t]
\begin{center}
\includegraphics{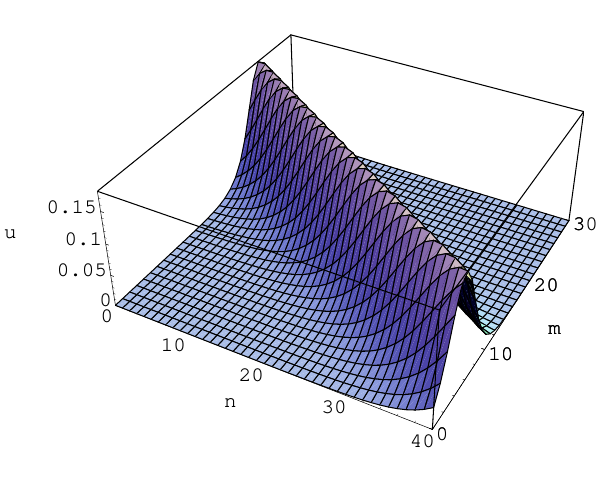}
\end{center}
\caption{Soliton at $a=1$, $b=2$, $k=0.3$, $p=q=x=0$, $\xi=-7.5$}
\label{fig:sol1}
\end{figure}

Let us start from the the simplest situation of constant parameters
$\alpha_m=\alpha$, $\beta_n=\beta$. In this case the linear seed solution is
obvious
\begin{equation}\label{seed}
 v_{m,n}=am+bn+px+q, \quad a^2-b^2=\alpha-\beta.
\end{equation}
Applying the B\"acklund transformation (\ref{KdVBT}) one obtains
\[
 \bar v_{m,n}=am+bn+px+q-k\tanh(kx+\phi_{m,n}), \quad \lambda=2p-k^2.
\]
Let us search for the dependence of $\phi$ on $m,n$ in the form
\[
 \phi_{m,n}=\mu m+\nu n+\xi
\]
and denote
\[
 e^{2(kx+\phi_{m,n})}=X,\quad e^{2\mu}=M,\quad e^{2\nu}=N.
\]
Then substitution into (\ref{homdpKdVmn}) yields
\[
 \left(a+b+k\frac{X-1}{X+1}-k\frac{MNX-1}{MNX+1}\right)
 \left(a-b+k\frac{NX-1}{NX+1}-k\frac{MX-1}{MX+1}\right)=a^2-b^2.
\]
Collecting together and splitting over $X$ give the relations between
the parameters which can written in the form
\[
 a\tanh\mu=b\tanh\nu=k,
\]
so that finally 1-kink solution is given by the formulae
\begin{equation}\label{kink}
 v_{m,n}=am+bn+px+q-k\tanh(kx+\mu m+\nu n+\xi),
\end{equation}
where
\[
 a^2-b^2=\alpha-\beta, \quad
 \mu=\frac{1}{2}\log\frac{a+k}{a-k}, \quad
 \nu=\frac{1}{2}\log\frac{b+k}{b-k}.
\]
Obviously, one has to choose $|k|<\max\{|a|,|b|\}$ in order to obtain real
$\mu,\nu$. The dependence on the parameter $x$ can be neglected, but we may
also make use of it in order to obtain the discrete dynamics of the soliton
solution $u_{m,n}=-2\partial_x(v_{m,n})$ which corresponds to the usual KdV
equation $u_t=u_{xxx}+6uu_x$ (see Fig.~\ref{fig:sol1}).

One can check using the formula (\ref{KdVBT}) that the corresponding fields $u$
satisfy the following {\em discrete KdV equation}
\begin{equation}
\label{dKdV'}
   \sqrt{u_{12}+u_2-2\alpha}-\sqrt{u_1+u-2\alpha}
  =\sqrt{u_{12}+u_1-2\beta}-\sqrt{u_2+u-2\beta}
\end{equation}
or, equivalently,
\begin{multline}
\label{dKdV}
 \frac14(u_{12}-u)^2(u_1-u_2)^2-(\alpha-\beta)(u_{12}-u)(u_1-u_2)(u+u_1+u_2+u_{12})\\
 +(\alpha-\beta)^2((u_{12}-u)^2+(u_1-u_2)^2)
 +2(\alpha^2-\beta^2)(u_{12}-u)(u_1-u_2)=0.
\end{multline}
In contrast to the discrete potential KdV equation this equation is  quadratic
with respect to each field, and thus is not covered by the classification
\cite{ABS}.

Two-kink solution can be obtained via NSP for B\"acklund
transformation, which brings to the formula
\[
 v^{(12)}=v^{(0)}
 -\frac{k^2_2-k^2_1}{k_2\tanh(k_2x+\phi_2)-k_1\tanh(k_1x+\phi_1)}
\]
where the seed solution $v^{(0)}$ and phases $\phi_1$, $\phi_2$ are defined as
above. In order to obtain profile without singularity in $x$ one may choose
$k_2>k_1>0$ and make a shift $\phi_2\to\phi_2+\frac{\pi i}2$. In this way the
general multikink solution can be constructed. Notice the different effect of
the B\"acklund transform which may result either in the dressing of the
solution or just in the phase shifts $m\to m+1$, $n\to n+1$.

\begin{figure}[ht]
\centerline{\includegraphics{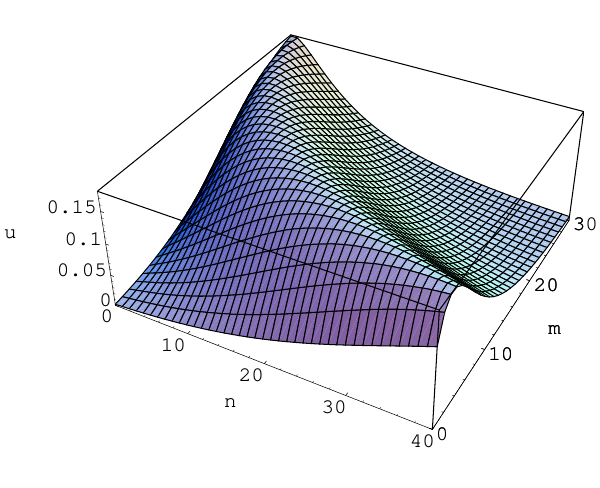}}
\caption{Bended soliton}\label{fig:solbend}
\end{figure}

The seed solution (\ref{seed}) is easily generalized for the case of variable
parameters:
\[
 v_{m,n}=a_m+b_n+px+q,\quad
 (a_{m+1}-a_m)^2-\alpha_m=(b_{n+1}-b_n)^2-\beta_n=\delta.
\]
(Notice that this is more general ansatz even in the case of constant
$\alpha,\beta$.) Proceeding as above one obtains the kink moving with the
variable velocity:
\begin{equation}\label{kink2}
 v_{m,n}=a_m+b_n+px+q-k\tanh(kx+\mu_m+\nu_n+\xi),
\end{equation}
where
\[
   (a_{m+1}-a_m)\tanh(\mu_{m+1}-\mu_m)
  =(b_{n+1}-b_n)\tanh(\nu_{n+1}-\nu_n)=k,
\]
that is
\[
 \mu_{m+1}-\mu_m=\frac{1}{2}\log\frac{a_{m+1}-a_m+k}{a_{m+1}-a_m-k},\quad
 \nu_{n+1}-\nu_n=\frac{1}{2}\log\frac{b_{n+1}-b_n+k}{b_{n+1}-b_n-k}.
\]
On the Fig.~\ref{fig:solbend} the graph of the soliton $u=-2v_x$ is plotted
corresponding to the values
\begin{gather*}
 a_0=b_0=\mu_0=\nu_0=p=q=x=0,\\
 a_m-a_{m-1}=\sqrt{m},\quad b_n-b_{n-1}=\sqrt{2n},\quad
 k=0.3,\quad \xi=-3.
\end{gather*}

Moreover, the formula (\ref{kink2}) can be easily generalized for the case of
multidimensional lattice $\Integer^d$ in which any 2D sublattice is  governed
by dpKdV equation:
\begin{equation}\label{kinkd}
 v_{n_1,\dots,n_d}=a^1_{n_1}+\dots+a^d_{n_d}+px+q
 -k\tanh(kx+\nu^1_{n_1}+\dots+\nu^d_{n_d}+\xi),
\end{equation}
where
\begin{gather*}
 (a^i_{n+1}-a^i_n)^2-\alpha^i_n=\delta,\quad
 (a^i_{n+1}-a^i_n)\tanh(\nu^i_{n+1}-\nu^i_n)=k, \\
 \nu^i_{n+1}-\nu^i_n=\frac{1}{2}\log\frac{a^i_{n+1}-a^i_n+k}{a^i_{n+1}-a^i_n-k}.
\end{gather*}
This observation immediately gives us the kink solution for the quad-graphs
which can be embedded into $\Integer^d$ lattice. For instance, the graph shown
on the Fig.~\ref{fig:qg}d can be obtained from the cubic lattice by taking all
its faces which intersect the plane $n_1+n_2+n_3=0$, so that the kink solution
for this graph is given by the formula (\ref{kinkd}) at $d=3$ restricted on the
values $n_1+n_2+n_3\in\{-1,0,1\}$. In the case of the linear phase
$\nu^i_{n_i}=\nu^in_i$ the corresponding soliton propagates in the direction
$(\nu^3-\nu^2, \nu^1-\nu^3,\nu^2-\nu^1)$ (see Fig.~\ref{fig:solhex}).

\newcommand{\dgr}[2]{\centerline{\raisebox{40mm}{\includegraphics[width=30mm]{#1}}
\hspace{5mm}\includegraphics[width=85mm]{#2}}}

\begin{figure}[htbp]
\dgr{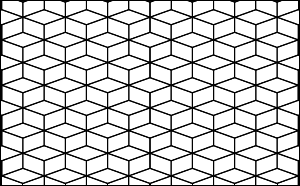}{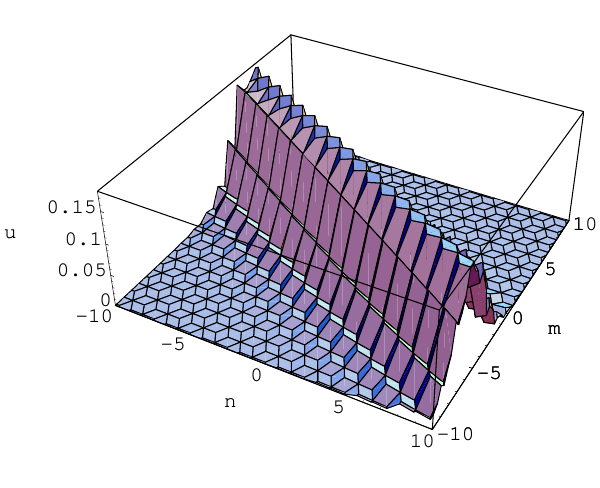}
\caption{Soliton on the regular quad-graph
($a_1=0.5$, $a_2=2.5$, $a_3=3$, $k=0.3$, $p=q=x=\xi=0$)}
\label{fig:solhex}
\end{figure}

\begin{figure}[htbp]
\dgr{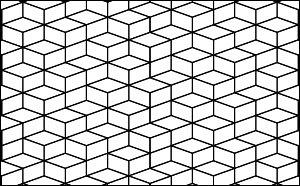}{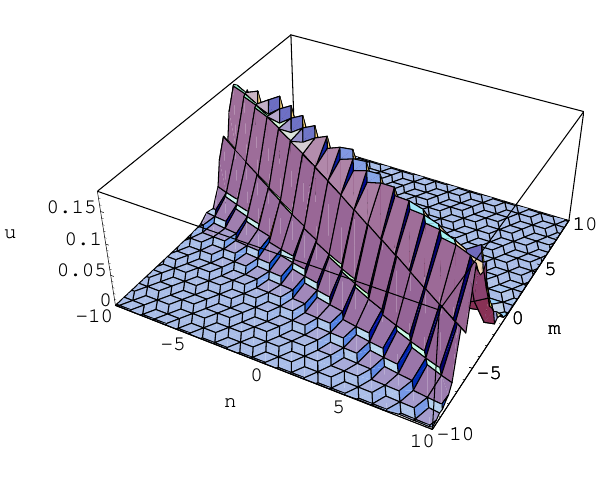}
\caption{Soliton on the quasiregular quad-graph}\label{fig:solquasihex}
\end{figure}

Some subclass of the quasiperiodic tilings can be obtained if one choose the
other planes and use de Bruijn's projection method (see e.g. \cite{Sen}). The
formula (\ref{kinkd}) works for such quad-graphs as well. On the
Fig.~\ref{fig:solquasihex} the same 3D solution is presented, restricted on the
plane $n_1+\sqrt{2}n_2+\sqrt{3}n_3=0$.

\begin{figure}[htbp]
\dgr{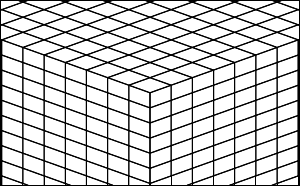}{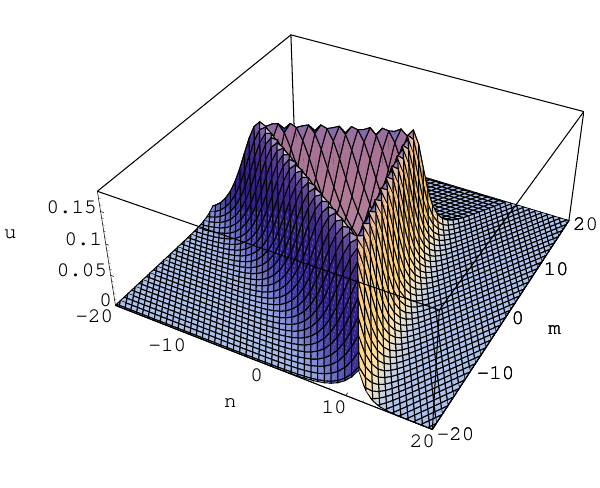}
\caption{Soliton on three quadrants $n_i>0$
($a_1=1$, $a_2=1.5$, $a_3=2$, $k=0.3$, $p=q=x=0$, $\xi=-3$)}
\label{fig:solquadrants3}
\end{figure}

\begin{figure}[htbp]
\dgr{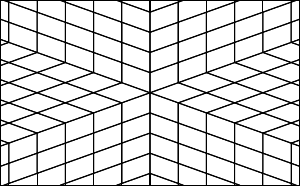}{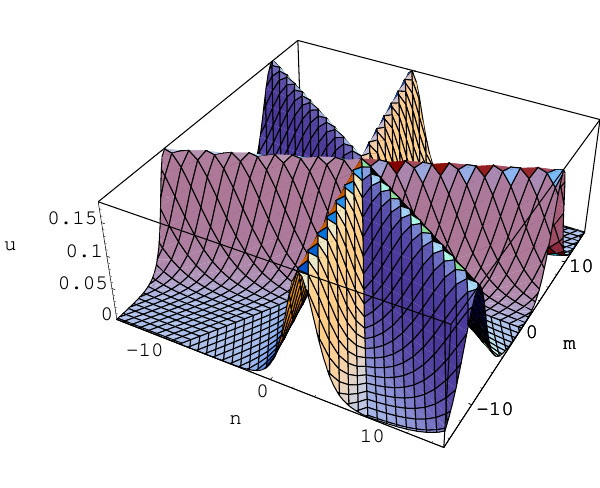}
\caption{Soliton on six quadrants $n_in_j<0$
($a_1=1$, $a_2=1.5$, $a_3=2$, $k=0.3$, $p=q=x=\xi=0$)}
\label{fig:solquadrants6}
\end{figure}

Two more figures Fig.~\ref{fig:solquadrants3} and \ref{fig:solquadrants6}
present the 3-dimensional solution restricted on the coordinate quadrants.

An interesting problem is to investigate the interaction of the kink  solutions
with the localized defects in the square lattice. According to the Theorem
\ref{th:defect} a weak defect in the homogeneous lattice does not affect
solutions, so that the ordinary kink (\ref{kink}) does not change. In contrast,
it can be shown that the bended kink (\ref{kink2}) may acquire some phase shift
after the interaction with such a defect.

\section{Discrete linear wave equation}\label{s:lin}

It is interesting to compare the theory of the homogeneous dpKdV equation we
discussed above with the discrete linear wave equation
\begin{equation}\label{wave}
 v_{12}-v_1-v_2+v=0
\end{equation}
on quad-graphs.

Obviously, equation (\ref{wave}) always can be solved with respect to any
vertex and it does not contain singularities in contrast to the dpKdV case.
Therefore the splitting of self-intersecting strips is not valid in this case.

The most serious distinction is that since the discrete wave equation does not
contain the parameters the corresponding matrices (\ref{ZCR})
\[
 L_{j,i}=\begin{pmatrix}
  1 & v_j-v_i \\
  0 &    1
 \end{pmatrix}
\]
contain no spectral parameter and give the fake zero curvature representation.
In particular, the refactorization scheme from the Section \ref{s:LL} does not
work, and the Theorem \ref{th:defect} is not valid, as the following simple
examples demonstrate. Of course, for the wave equation it is sufficient to
analyze the propagation of delta-like initial data (black circle denotes 1,
white one denotes $-1$, no circle denotes 0).

\def\QG{
\path(0,20)(105,20)
\path(0,40)(20,40)(40,30)(80,40)(105,40) \path(20,40)(60,50)(80,40)
\path(0,60)(105,60) \path(0,80)(105,80)
\path(20,0)(20,85)\path(40,0)(40,30)
\path(60,0)(60,35)(40,45)(40,85)\path(60,50)(60,85)
\path(80,0)(80,85)\path(100,0)(100,85)
\bfput{\path(0,85)(0,0)(105,0)}}
\begin{figure}[htb]
\begin{center}\setlength{\unitlength}{0.07em}
\begin{picture}(130,110)(-30,-10)
 \QG \put(20,-20){a}
 \multiput(20,0)(0,20){5}{\circle*{5}}
 \put(40,45){\circle*{5}}\put(40,60){\circle*{5}} \put(40,80){\circle*{5}}
\end{picture}
\begin{picture}(130,110)(-30,-10)
 \QG \put(20,-20){b}
 \put(40,0){\circle*{5}}\put(40,20){\circle*{5}}\put(40,30){\circle*{5}}
 \put(40,45){\circle{5}}\put(40,60){\circle{5}} \put(40,80){\circle{5}}
 \put(60,50){\circle{5}}\put(60,60){\circle{5}} \put(60,80){\circle{5}}
\end{picture}
\begin{picture}(130,110)(-30,-10)
 \QG \put(20,-20){c}
 \put(60, 0){\circle*{5}}\put(60,20){\circle*{5}}\put(60,35){\circle*{5}}
 \put(40,45){\circle*{5}}\put(40,60){\circle*{5}}\put(40,80){\circle*{5}}
\end{picture}
\end{center}
\caption{Interaction of the linear wave with the weak defect: splitting,
reflection, bending}\label{fig:weak}
\end{figure}
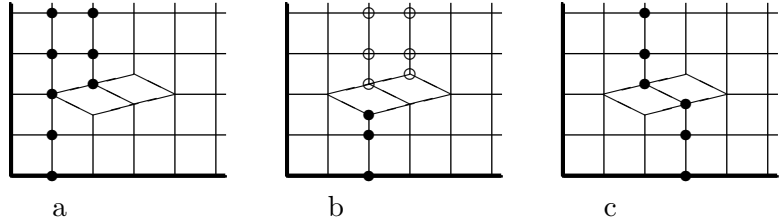

Nevertheless we can claim that a weaker version of the theorem \ref{th:defect}
is valid also for the discrete wave equation. Let us consider the Cauchy
problem for this equation on a regular square lattice with a weak defect $D$
with initial data on a staircase (or coordinate half-axes) outside the defect.
Again as in dpKdV case according to our Theorem \ref{th:corr} the solution
exists and unique and one can compare this solution with the one on the regular
lattice without defects.

Let us consider the {\em cross} $C(D)$ corresponding to the defect $D$ as the
union of all the characterstics passing through $D$. Besides the defect itself
it contains also all the vertical and horizontal strips passing through the
rectangle.

\begin{theorem}\label{th:wave}
For the discrete wave equation a weak defect $D$ does not affect the
solution outside the cross $C(D)$.
\end{theorem}

The proof easily follows from the fact that for the discrete wave equation the
difference $v-v_1$ is preserved when we are moving along the corresponding
charactersitic. Actually the affected area is only a half of this cross
(``thick hook'' as on Fig. \hyperref[fig:G_shade]{\ref*{fig:G_shade}b}). For a
general equation one should expect the whole quadrant bounded by this hook to
be affected.

\section{Concluding remarks}\label{s:conc}

We have shown that for the integrable equations on the quad-graphs it is
possible to formulate a criterion for the well-posedness of the Cauchy
problems. The 3D consistency condition plays a crucial role in our analysis.
The existence of Lax matrices provides an effective integration scheme for the
corresponding IVP problems based on matrix factorization problem.

The question is whether such a criterion exists for a general equation on
quad-graphs. One can easily produce a lot of examples of graphs and IVP which
are not sensitive to a particular form of the equation, but to describe all of
them in some geometric way seems to be a hard problem.

We have not discussed also global geometrical aspects (e.g. the quad-graphs on
the surfaces of genus $g$). For some interesting results in this direction in
the linear case we refer to Novikov and Dynnikov papers \cite{ND,No,DN,DN02}.

The discrete equations in three-dimensions in the context of the classical
geometry are investigated in the recent very interesting papers
\cite{KW1,KW2,KingSchief} by B. Konopelchenko, W. Schief and A. King. In
particular in \cite{KingSchief} an example of a well-posed Cauchy problem for
the discrete BKP equation is discussed.

\paragraph{Acknowledgments.}  The research by V.A. was partly supported by the
Alexander von Humboldt Stiftung and by the RFBR grant 02-01-00144.

A.V. is grateful to S.P. Novikov who stimulated his interest to the problems on
the graphs.


\end{document}